\documentclass[twocolumn,showpacs,amsmath,amssymb,superscriptaddress,nofootinbib]{revtex4}

\usepackage{graphicx}
\usepackage{epsfig}
\usepackage{dcolumn}
\usepackage{bm}

\begin{document}

\title{Conservation Properties in the Time-Dependent Hartree Fock
Theory}
\author{Lu Guo\footnote{Present address: Department of Physics, University of Tokyo, Hongo,
Tokyo 113-0033, Japan}}
\author{J. A. Maruhn}
\affiliation{Institut f\"ur Theoretische Physik, Universit\"at
Frankfurt, 60438 Frankfurt, Germany}
\author{P.--G. Reinhard}
\affiliation{Institut f{\"u}r Theoretische Physik $\mathit{II}$,
Universit{\"a}t Erlangen-N{\"u}rnberg, 91058 Erlangen, Germany}
\author{Y. Hashimoto}
\affiliation{Graduate School of Pure and Applied Sciences,
University of Tsukuba, Tsukuba 305-8571, Japan}

\begin{abstract}
We discuss the conservation of angular momentum in nuclear
time-dependent Hartree-Fock calculations for a numerical
representation of wave functions and potentials on a
three-dimensional cartesian grid. Free rotation of a deformed
nucleus performs extremely well even for relatively coarse spatial
grids. Heavy ion collisions produce a highly excited compound system
associated with substantial nucleon emission. These emitted nucleons
reach the bounds of the numerical box which leads to a decrease of
angular momentum.  We discuss strategies to distinguish the
physically justified loss from numerical artifacts.
\end{abstract}

\maketitle


Time-dependent Hartree-Fock (TDHF), originally proposed by
Dirac~\cite{Dirac-tdhf}, has found widespread applications in various
areas of physics due to the overwhelming development of computational
power.  It is employed, e.g., as the variant time-dependent density
functional theory \cite{Gro90} in atomic, molecular and cluster
physics, see e.g. \cite{Rei03a,tddft-notes}. It has enjoyed application in
nuclear dynamics since more than thirty years~\cite{Bonche76} as a
microscopic approach to various dynamical scenarios in the regime of
large amplitude collective motion, like fusion excitation functions, fission,
deep-inelastic scattering, and collective excitations; for early reviews
see, e.g., \cite{Negele82,Davies85}.
With the steady upgrade of computational power, three-dimensional TDHF
calculations without any symmetry restriction became possible and
renewed the interest in nuclear TDHF as seen from an impressive series
of recent publications
\cite{Simenel03,Simenel04,Naka05,Umar05a,Maruhn05,Maruhn06,Umar06a,Umar06b,Guo07a,Guo07}.
A crucial aspect in nuclear TDHF is that nuclei are freely moving
objects such that all conservation laws (energy, momentum, angular
momentum) apply. Conservation of energy and momentum is a basic
feature which has been tested for all existing codes. Conservation of
angular momentum has not yet been studied and that is the topic which
we want to address in this paper.

\begin{figure}
\centerline{\epsfig{figure=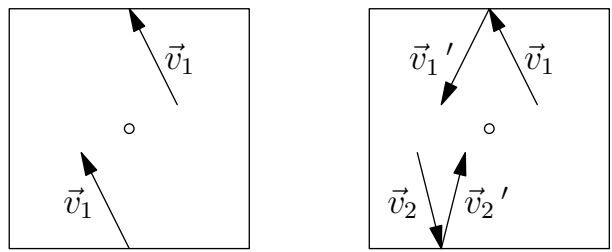,width=8cm}}
\caption{\label{fig:boundary}
Simple illustation of the effect of boundary conditions on total
angular momentum. The boxes indicate the computational boundaries and
the central dot the reference point for the angular momentum. For
periodic boundary conditions (left) it is easy to even revert the sign
of the particle's angular momentum. For approximately reflecting
boundary conditions (right) the situation is not quite as pronounced:
for the case $\vec v_1$ there is no change by reflection, while for
$\vec v_2$ the sign also changes.}
\end{figure}

The calculations employ grids in coordinate space. Their finite
spacing and box size destroy translational and rotational symmetry
which, in turn, can spoil conservation of momentum and angular
momentum. The major destructive mechanism comes from matter which
tries to leave the computational box but is hindered by the boundary
condition. In the course of time development, higher energy
components appear in the nucleon wave functions, representing
outgoing ``particles'' from the nucleus. The further time
development may therefore be affected by their reflection from the
boundary or, in the case of periodic boundary conditions, re-entry
from the neighboring cells.  That can change the total angular
momentum, as is illustrated in Fig.~\ref{fig:boundary}. Clearly the
boundary can even change the sign of a particle's contribution to
the angular momentum around the center of the cell.
In this work we consider periodic boundary conditions. The case of
reflecting boundaries behaves qualitatively similar.

The static and time-dependent Hartree-Fock equations are solved on a
cartesian three-dimensional mesh without any symmetry requirements.
The grid spacing was 1~fm, and the Skyrme energy functional
\cite{Bender03} was employed with the parametrization SLy6
\cite{Chabanat98} (for the purposes of this work, the particular
choice of Skyrme force is irrelevant). The minimum set of time-odd
terms to assure Galilei invariance \cite{{Bender03,Eng75a,Doba95}} 
was included. The spatial derivatives are
calculated using the Fast Fourier Transform and periodic boundary
conditions are employed, except for the Coulomb potential, which is
calculated with boundary conditions at infinity as described in
\cite{Eas79}. The time stepping employs a sixth-order Taylor
expansion of the time evolution operator $U(t,t+\Delta t)=
\exp\left[-{\mathrm{i}h(t+\Delta t/2)\,\Delta t}/\hbar\right], $
with the mean fields at the half step estimated by a third-order
expansion using the mean field at time $t$.

This method of time development is non-unitary, so that the
orthonormality of the single-particle wave functions is not
guaranteed. Nevertheless, we find that the calculation is quite
stable and accurate for several thousand time steps $\Delta t\approx
0.2{\rm~fm}/c$, in the sense that the particle number changes by
less than 0.1\%. The total energy also is conserved to a fraction of
an MeV.  When instability then sets in, there is a rapid drift of
particle number and energy, so that the conservation properties are
quite good checks for the accuracy and stability of the calculation.

Test case will be a collision of two $^{16}$O nuclei which involves
truly large amplitudes and carries a large amount of excitation
energy.
As argued above, the boundaries of the numerical grid influence the
conservation laws. In order to disentangle its effects, we consider
three different setups:
\begin{enumerate}
\item\label{it:small}
A small grid with $24\times32\times32\, {\rm fm}^3$.
\item
A doubled grid of $48\times64\times64\,{\rm fm}^3$,
so that the collision is surrounded more generously by empty space.
Observables are summed over only the smaller grid of case
(\ref{it:small}), which allows to distinguish the exact physical loss
from the artifacts of the smaller grid.
\item
Finally, the small grid plus an absorbing boundary conditions which
are arranged within an absorbing layer $N_{\rm abs}=6$ cells wide in
each direction. In this layer, a mask function
$M(n_x,n_y,n_z)=M_x(n_x)M_y(n_y)M_z(n_z)$ is applied to the wave
functions after each time step where $ M_i(n_i)=\rm cos
\left((N_{\rm abs}\!+\!1\!-\!n_i){\pi}/ {2}{N_{\rm
abs}}\right)^{0.25}\,,\ n_i=1\ldots N_{\rm abs} $. See
\cite{Rein06} for details.
\end{enumerate}


Before analyzing rotational motion, we have checked, of course the
conservation of energy and total momentum. Both quantities are
conserved very well with relative fluctuations staying at the order of
10$^{-4}$.  A detailed analysis of translational motion and of the
physical interpretation of the TDHF single-particle energies can be
found in \cite{Guo07}. Let us just mention here that for a nucleus
moving freely in any direction on the grid the momentum is conserved
to an accuracy of better than $10^{-4}$.

For the case of angular momentum, the situation turned out to depend
strongly on the excitation of the system. We therefore discuss two
types of calculations: single cranked nucleus and heavy-ion
collisions.

\subsection{Single cranked nucleus}

We produce a rotating nucleus by solving the cranked static
Hartree-Fock equations
\begin{equation}
  \left(\hat h-\omega{\hat J}_x\right)\,\phi_k(\vec r)
  =
  \varepsilon_k\phi_k(\vec r),\qquad k=1\ldots A
\end{equation}
with $\omega$ the prescribed angular frequency of rotation about the
$x$-axis (for simplicity we omit spin dependence, though spin is
included in the calculations). The rotating states
\begin{equation}
  \bar\phi_k(\vec r,t)
  =
  \exp\left[-\frac{{\rm i}\omega t \hat J_x}\hbar\right]
  \;\exp\left[-\frac{{\rm i}\varepsilon_k t}\hbar\right]\;\phi_k(\vec r)
\end{equation}
then are exact solutions of the TDHF equations, so that the numerical
solution should show simply a rotating nucleus with no extraneous
motion added.

It is clear that this is a much more demanding test for the numerical
solution, since the cartesian grid is incompatible with rotational
motion and effectively the grid spacing expands and shrinks by a
factor of $\sqrt2$ as the nucleus rotates through 45 and then 90
degrees.

\begin{figure}
\centerline{\epsfig{figure=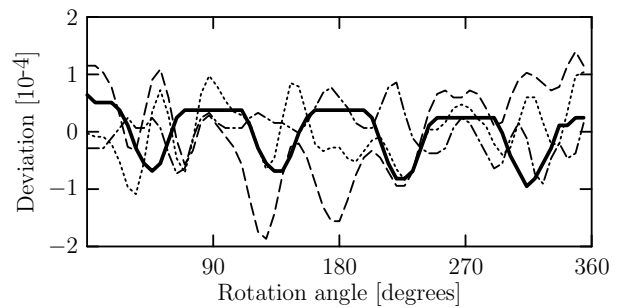,width=8cm}}
\caption{\label{fig:rotate} Relative deviations of the angular
momentum expectation value (full curve) and the three principal
moments of inertia (smallest: dotted, intermediate: dashed, and
> largest: dot-dashed curve) from the temporal average during rotation of a
$^{24}$Mg nucleus. The abscissa denotes the rotation angle
calculated from the instantaneous tensor of inertia.}
\end{figure}

As an example, we show here the deformed nucleus $^{24}$Mg cranked with
$\omega=2\,{\rm MeV}/\hbar$. In this case, rotation turns out to be
almost completely of rigid-body type; the observed angular momentum of
$7.54\,\hbar$ corresponds to a moment of inertia of about 98\% of the
rigid-body value.  To judge the accuracy of the rotation in the
numerical solution, we examine the expectation value of the angular
momentum ${\hat J}_x$, which should be strictly conserved, and the
principal moments of inertia, which in the absence of any internal
excitation should also be constant. Fig.~\ref{fig:rotate} shows the
fluctuations of these quantities as functions of simulation time.

The most striking result is the excellent quantitative description of
rotation in spite of the coarse Cartesian grid, the variations being of
the order of $10^{-4}$. The angular-momentum expectation value clearly
is correlated with the angle and shows regular variations with a 45°
structure. The variations in the moments of inertia show a less
regular pattern and appear to be influenced by the periods of the
internal vibrations of the nucleus from which seemingly a tiny amount
is exited during the rough sliding over the grid.

\subsection{Heavy-ion reactions}

\begin{figure*}
\centerline{\epsfig{figure=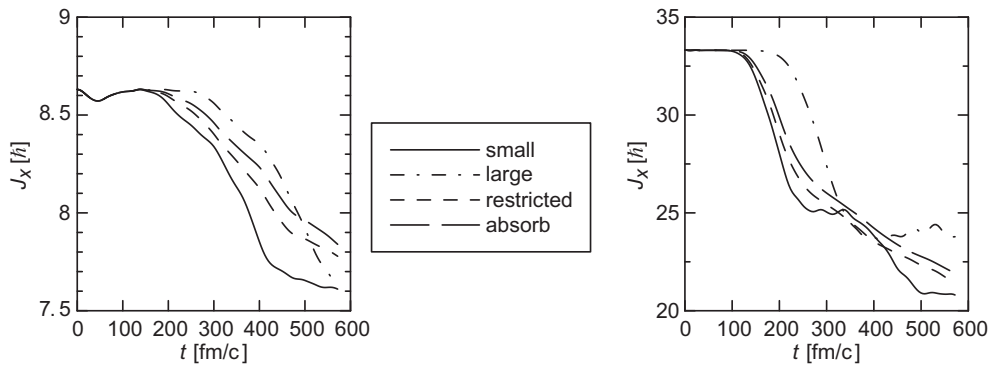,width=13cm}}
\caption{\label{fig:coll} Angular momentum as a function of
time for collisions of $^{16}$O+$^{16}$O at $E_{\rm cm}=25\,$MeV
(left) and $125\,$MeV (right). The different cases are: ``small'':
calculation in a grid of $24\times32\times32\,{\rm fm}^3$,
``large'': calculation in a grid doubled in total size in every
direction, ``restricted'': same as large, but angular momentum is
summed up only over the small grid; ``absorb'': small grid with a
6~fm absorbing layer around the boundary.}
\end{figure*}

The situation is quite different in the more violent case of
heavy-ion collisions. The substantial excitation leads to emission
of nucleons which, in turn, causes problems.
As was shown in the introduction, particles crossing the boundaries
can generate large spurious changes in the total angular momentum
and it becomes quite difficult to separate the correct physical loss
of angular momentum carried away by the emitted particles from the
spurious numerical effect. The results presented in
Fig.~\ref{fig:coll} illustrate the problems which are of quite
general nature. Note that the angular momentum $J_x$ 
is perpendicular to the reaction plane.

The initial condition consists of two ground-state $^{16}$O nuclei
with a c.m.\ energy of either 25 or 125~MeV. All the initial angular
momentum thus comes from the relative motion. For the higher energy
the impact parameter was $b=4.8$~fm, corresponding to
$J_x\approx33\hbar$, while for the lower energy $b=2.8$~fm
corresponding to $J_x\approx8.6\hbar$. These values were chosen 
to have the same distance of closest approach in the pure Coulomb trajectory.

For both energies, the two nuclei stay fused at the end of the
calculation. The boundary problems are more serious for the higher
energy because of the higher excitation leading to stronger emission
of particles. It is apparent that in the larger grid the reduction
in angular momentum starts about $100\,{\rm fm}/c$ later as compared
to the small grid. The curve labeled ``restricted'' is computed in
the large grid while the angular momentum is collected in the small
grid. This should indicate the true loss of angular momentum from
the small grid for the time span before emitted particles come back
from the larger boundary. Clearly the ``small'' calculation has the
largest loss and it becomes even unreasonable at about $250\,{\rm
fm}/c$. It is reassuring that the curve for the absorbing boundary
stays quite close to the ``restricted'' calculation and shows a
reasonable monotonic decrease throughout.

The total reduction of about 30\% at a collision energy of 125 MeV
is surprisingly large in view of the fact that only 1.7 nucleons are
absorbed. These nucleons thus carry a comparatively large share of
angular momentum.
At the lower energy of $25\,$MeV, the total change in angular momentum
is not as dramatic but by no means negligible, still exceeding 10\%
while 0.4 nucleons are emitted.  Besides the quantitative difference,
the general pattern are very similar.  Again, the loss sets in later
for the larger grid and the calculation with the absorbing layer
appears to be a reasonable approximation to the ``true'' loss.


In this paper, we have analyzed the conservation of total angular
momentum in nuclear TDHF calculations. The calculations used a
coordinate-space representation of wave functions and potential
fields on a three-dimensional cartesian grid without any symmetry
restriction.  The full Skyrme interaction was taken into account.
Conservation of energy and momentum was tested (but not detailed
here) and found to be well matched within a relative error of only
10$^{-4}$.

The results for angular momentum depend on the dynamical scenario.
Free rotation of a deformed nucleus is surprisingly well described.
Although the cartesian grid spoils rotational symmetry, we find that
the deformed nucleus rotates steadily over the grid of 1 fm spacing
with variations of angular momentum and moments of inertia of the
order of 10$^{-4}$. This is the same quality as found already for
translational momentum and energy.

The case of nucleus-nucleus collisions is less well-behaved.  The
compound system is heavily excited. This leads to substantial emission
of nucleons which in the sequel reach the bounds of the numerical box
where reflection or periodic copy (depending in the grid model) lead
to a substantial reduction of angular momentum.  Comparing
calculations on different grids (small box, large box, absorbing
bounds), we have worked out that the loss is to a large extent
physical because the emitted nucleons are very energetic and carry
away a comparatively large amount of angular momentum. Artifacts from
the boundary come into play as soon as nucleons travel back into the
reaction zone. This happens the later the larger the grid and it can
be effectively avoided when using absorbing boundary conditions. Both
``solutions'', larger or absorbing grid are somewhat expensive. One
may live with a small grid if one confines the analysis to the early
time evolution of the compound system (to evaluate the doorway
effects).  In any case, the conserved quantities should be checked
carefully for each new dynamical scenario.

The results discussed here are, of course, only indicative and may
vary quantitatively for other TDHF codes. The specific discretization
of the equations of motion will affect the accuracy of describing an
isolated rotating nucleus, while the treatment of the boundary
conditions will strongly influence the boundary problems addressed
above.

Lu Guo acknowledges support from the Alexander von Humboldt Foundation.
We gratefully acknowledge support by the Frankfurt Center for
Scientific Computing and from the BMBF (contracts 06 ER 124 and
06 F 131).

\bibliography{angmom2}

\end{document}